\documentclass[iop,onecolumn]{emulateapj}

\shorttitle{}
\shortauthors{Fujii, Okuzumi, \& Inutsuka}

\begin{document}

\title{A Fast and Accurate Calculation Scheme for Ionization Degrees in 
Protoplanetary and Circumplanetary Disks with Charged Dust Grains}

\author{Yuri I. Fujii\altaffilmark{}, Satoshi Okuzumi\altaffilmark{},
and Shu-ichiro Inutsuka\altaffilmark{}}
\affil{Department of Physics, Nagoya University, Furo-cho, Chikusa-ku,
Nagoya, Aichi 464-8602, Japan}
\email{yuri.f@nagoya-u.jp}

\begin{abstract}
We develop a fast and accurate calculation method for ionization degrees
in protoplanetary and circumplanetary disks including dust grains.
We apply our method to calculate the ionization degree of circumplanetary
disks.  It is important to understand the structure and evolution of 
protoplanetary/circumplanetary disks since they are 
thought to be the sites of planet/satellite formation. 
The turbulence that causes gas accretion is supposed to be driven by 
magnetorotational instability (MRI) that occurs only when the ionization 
degree is high enough for magnetic field to be coupled to gas. 
We calculate the ionization degrees in circumplanetary disks to
estimate the sizes of MRI-inactive regions.
We properly include the effect of dust grains because they efficiently 
capture charged particles and make ionization degree lower. 
Inclusion of dust grains complicates the reaction equations and requires 
expensive computation. In order to accelerate the calculation of 
ionization reactions, we develop a semianalytic method based on 
the charge distribution model proposed previously.
This method enables us to study the ionization state of disks for a wide range
of model parameters. For a previous model of circum-Jovian disk,
we find that an MRI-inactive region covers almost all regions even without dust grains. 
This suggests that the gas accretion rates in circumplanetary disks 
are much smaller than previously thought.
\end{abstract}

\keywords{dust, extinction -- planets and satellites: formation -- protoplanetary disks}

\section{Introduction}

Various observations suggest substantial gas accretion disks around
young stellar objects, so-called protoplanetary disks. 
Transport of the angular momentum is needed for 
gas accretion. At present, the magnetohydrodynamic (MHD) turbulence driven by 
magnetorotational instability (MRI) is thought to be the most 
promising mechanism for angular momentum transport. 
However, protoplanetary disks have very low fraction of charged components.
This is due to their low temperature and high density; recombination 
is efficient in high density gas, and thermal 
ionization does not work except for very inner regions.
In order to understand disk evolution, we must
clarify which part of the disk is magnetorotationally unstable. 
Magnetic Reynolds number $Re_{\rm m}\equiv \rm UL/\eta$ 
(U and L are characteristic velocity and length respectively, and $\eta$ is
magnetic diffusion coefficient) is used to be a measure of such instability.
If $Re_{\rm m}$ is large, the region is a magnetically 
active zone; if $Re_{\rm m}$ is small, the region is 
a magnetically inactive, so-called, ``dead zone''\citep{gam96}.
Since $\eta$ is inversely proportional to the ionization degree, 
the investigation of ionization degree is important to estimate the 
value of $Re_{\rm m}$, or the location of the MRI-inactive region.

A number of studies on the ionization degree in protoplanetary disks 
have appeared in the literature (e.g., \citealt{san00}; \citealt{ilg06};
\citealt{oku09}).
These studies assume steady state for reactions and do not consider 
the time-dependent ionization events.
However, some observation has shown that young stars emit X-ray flares 
whose timescales are order of a day \citep{wol05}.
Some of the dynamical timescales (e.g., reconnection) are expected
to be very short.
These facts indicate that time-dependent calculation is needed to 
investigate the ionization degree in realistic dynamical environments.

However, it remains difficult to calculate highly 
time-dependent ionization degree numerically because the network of 
chemical reactions in the gas of the disks is highly complicated. 
Some studies on the ionization degree in protoplanetary disks 
(e.g. \citealt{fro02}) did not consider the effect of dust 
grains since the inclusion of their effect complexifies the reactions further;
but the effect of dust grains cannot been ignored, 
since very efficient capture of charged particles 
by dust grains makes the ionization degree much lower. 

There are many studies on the ionization degree of the protoplanetary 
disks, but no one has calculated that of circumplanetary disks 
with dust grains yet. It is important to understand the structure 
and evolution of circumplanetary disks, since the mass accretion through 
the disk onto the central planet is important in the early formation 
phases of the disks, and in addition, they might be the sites of 
satellites formation. 

In this work, we develop a fast and accurate time-dependent calculation 
method for the ionization degree in protoplanetary and circumplanetary 
disks. We try to reduce the computation time since we want
to plug our method into MHD simulations.

We adopt Gaussian distribution approximation for the charge distribution
of dust grains \citep{oku09}.
This approximation decreases the number of equations and allows us 
to calculate the ionization degree more efficiently especially when the dust 
grains have wide range of the charge distribution. 
We use the piecewise exact solution that is developed by \citet{ino08}.
In this method, we analytically solve some part of the reaction equation 
first, and use the solution as an initial condition of time integration 
of other terms. This solution enables us to calculate with larger time step. 
Our method can be applied to both circumplanetary disks and protoplanetary disks, 
and can be conveniently plugged into multi-dimensional MHD codes.


This paper is organized as follows. In Section 2, we describe the 
chemical reactions in the planetary disks, and derive equations for 
our calculation. The methods to speed up the calculation are shown. 
In Section 3, we test our method by calculating the ionization degree 
in the protoplanetary disks. In Section 4, we apply our method for 
circumplanetary disks, and discuss the occurrence of MRI.
Summery is in Section 5.
\section{Reactions and Equations}
\subsection{Reactions}
Gas in the protoplanetary and circumplanetary disks is mostly 
neutral hydrogen molecules.
However, they are thought to be ionized weakly 
by ionization source such as cosmic rays, X-rays, and ultraviolet radiation.
Resultant ionized particles make secondary ions and molecules, 
which enable further complex reactions.
We describe representative reactions. 

When hydrogen molecules are ionized 
\begin{equation}
 \rm   H_2 \ \longrightarrow H_2^+ + e,
\label{a}
\end{equation}
$\rm H_2^+$ immediately reacts with $\rm H_2$ to produce $\rm H^+_3$:
\begin{equation}
 \rm   H_2^+ + H_2 \ \longrightarrow H_3^+ + H.
\label{b}
\end{equation}
%
Reaction between $\rm H^+_3$ and molecules (e.g., CO) leads to heavier 
molecular ions:
\begin{equation}
 \rm   H_3^+ + CO \ \longrightarrow HCO^+ + H_2.
\label{c}
\end{equation}
Molecular ions are destroyed by charge exchange reactions 
with atomic heavy metals such as $\rm Mg$:
\begin{equation}
 \rm   HCO^+ + Mg \ \longrightarrow  \ Mg^+ + HCO.
\label{d}
\end{equation}
As long as atomic heavy metals are abundant, 
dissociative recombination
\begin{equation}
 \rm   HCO^+ + e^- \ \longrightarrow  \ CO + H,
\label{e}
\end{equation}
is slow, and reaction (Equation $(\ref{d})$) exceeds.
A small fraction of metal ions are destroyed by radiative recombination: 
\begin{equation}
 \rm Mg^+ + e^- \ \longrightarrow Mg + h\nu.
\label{f}
\end{equation}

We describe major molecular ions (e.g. $\rm H_2^+,\ H_3^+,\ and 
\ HCO^+$) as $\rm m^+$, 
and major heavy metal ions (e.g. $\rm Mg^+, Fe^+$) as $\rm M^+$ 
for simplicity (\citealt{opp74};
 \citealt{san00}; \citealt{ilg06}).

The effect of dust grains should be also considered. Ions 
and free electrons are captured by dust grains, and it lowers 
the ionization degree.
With sufficient dust grains, ionization degree is
determined mainly by electron capture rate of dust grains.
In Figure 1, we show the schematic view of reactions.

\begin{figure}[t]
\begin{tabular}{cc}
\begin{minipage}{0.5\textwidth}
\begin{center}
    \includegraphics[keepaspectratio, scale=0.5] {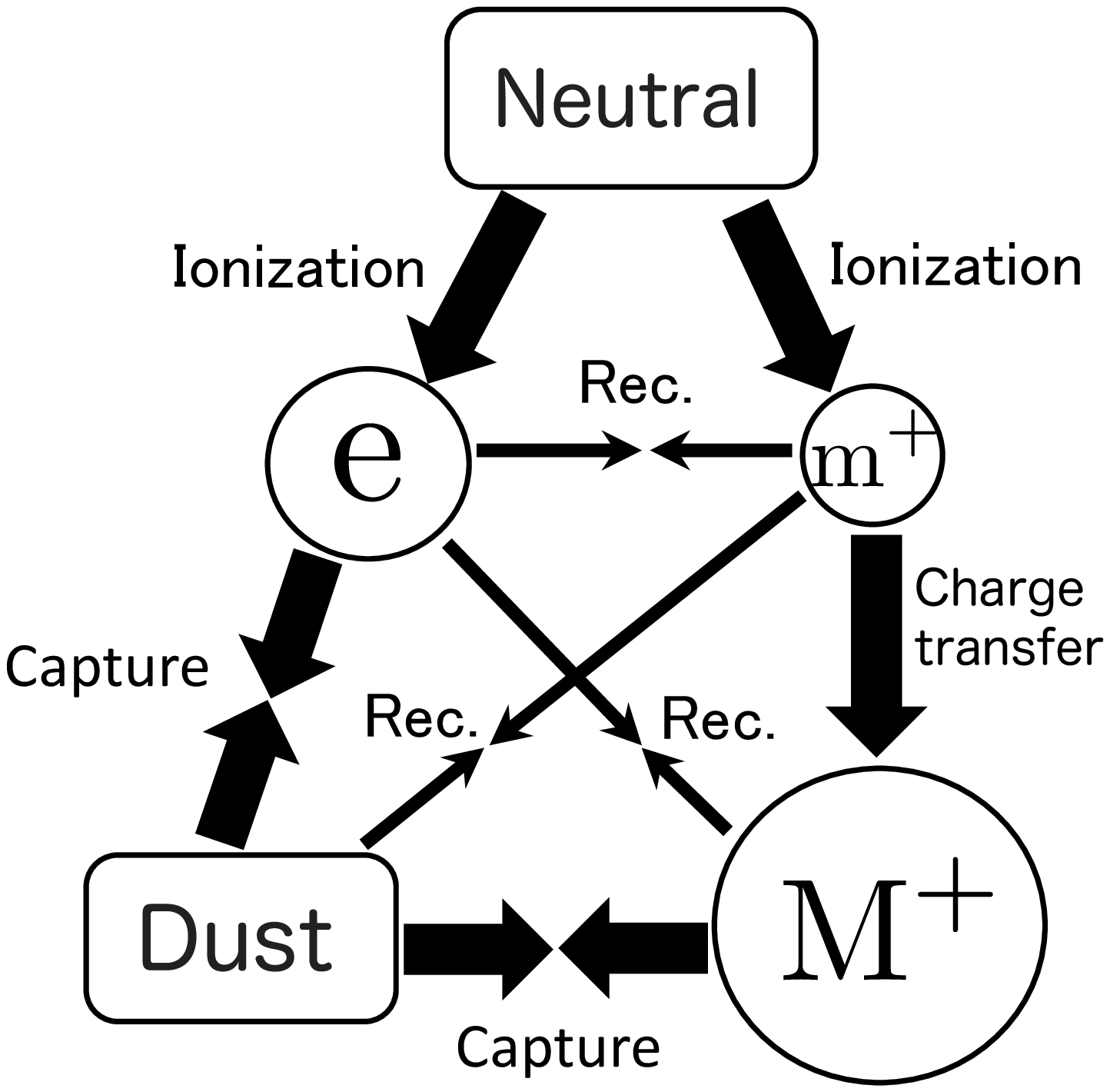}
    \caption{Schematic view of simplified reaction. 
    $\rm m^+,\ \rm M^+$, and e represent molecular ions, heavy metal
    ions, and electrons respectively. Neutral gas is ionized by the 
    ionization source, and $\rm m^+$ transfer their charge to $\rm M^+$.
    Ions and electrons are captured by dust grains, and 
    recombine a little.}
\end{center}    
\end{minipage}
\begin{minipage}{0.5\textwidth}
\begin{center}
    \includegraphics[keepaspectratio, scale=0.5] {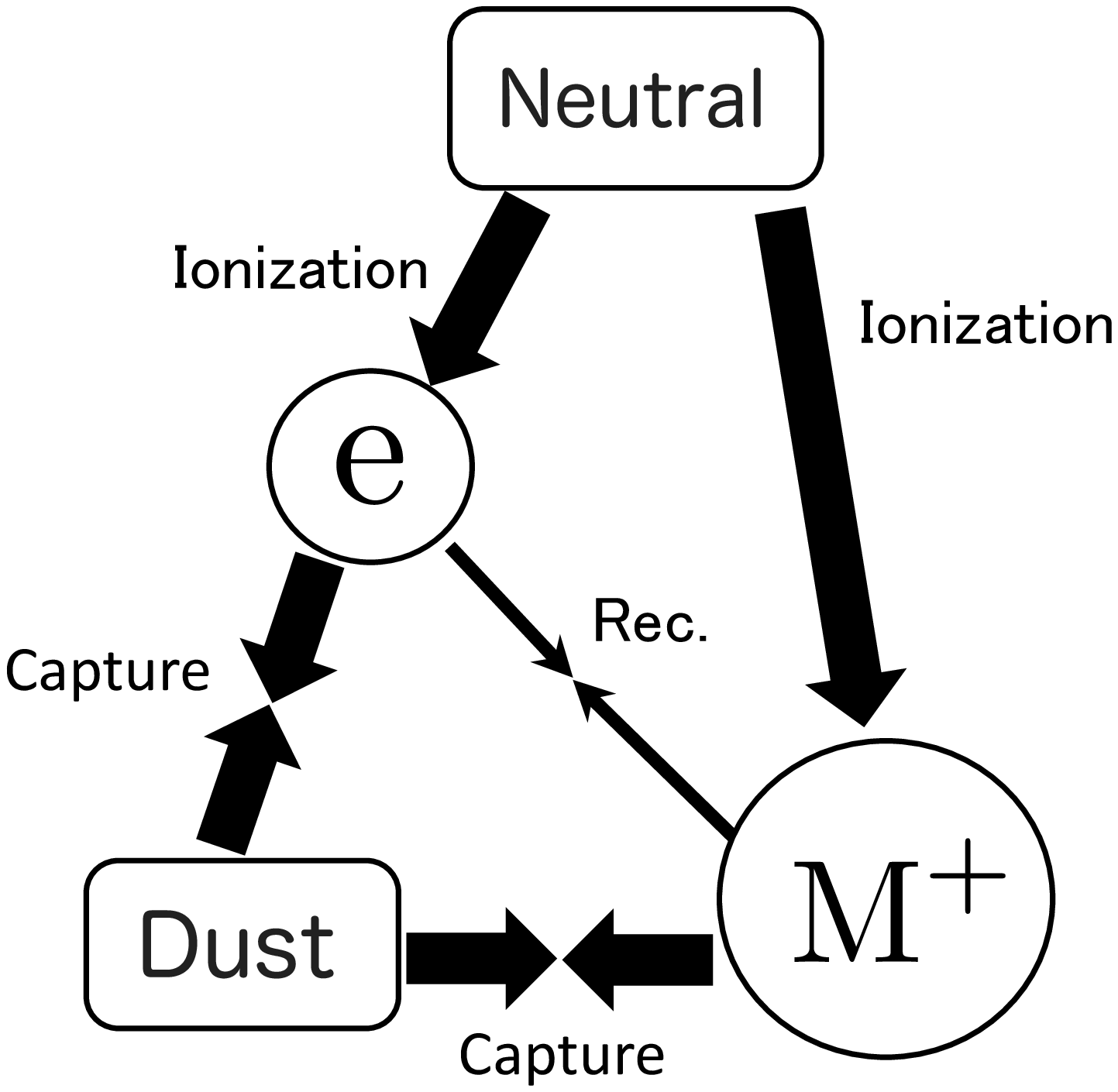}
    \caption{$\rm m^+$ is removed from Figure 1 because charge transfer
    is so fast that dissociative recombination and capture by dust grains 
    can be ignored.}
\end{center}    
\end{minipage}
\end{tabular}
\end{figure}

\subsection{Basic Equations}

Most of the gas in a protoplanetary disk is neutral,
so we assume that the number density of the neutral gas $n_{\rm n}$ 
is large enough and does not depend on ionization degree. 

The basic equations are the following:
\begin{equation}
    \frac{dn_{\rm m^+}}{dt} 
        = \zeta n_{\rm n} - \alpha_{\rm m^+}n_{\rm m^+}n_{\rm e} 
         - \beta n_{\rm m^+}n_{\rm M} 
         -\sum_{Z}^{}k_{\rm m^+ d}(Z)n_{\rm d}(Z) n_{\rm m^+},
    \label{221}
\end{equation}
\begin{equation}
    \frac{dn_{\rm M^+}}{dt} 
        = - \alpha_{\rm M^+}n_{\rm M^+}n_{\rm e} 
         + \beta n_{\rm m^+}n_{\rm M} 
         -\sum_{Z}^{}k_{\rm M^+ d}(Z)n_{\rm d}(Z) n_{\rm M^+},
    \label{222}
\end{equation}
\begin{equation}
    \frac{dn_{\rm e}}{dt} 
        = \zeta n_{\rm n} - \alpha_{\rm m^+}n_{\rm m^+}n_{\rm e} 
         - \alpha_{\rm M^+}n_{\rm M^+}n_{\rm e} 
       -\sum_{Z}^{}k_{\rm e d }(Z)n_{\rm d}(Z) n_{\rm e},
    \label{223}
\end{equation}
\begin{eqnarray}
    \frac{dn_{\rm d}(Z)}{dt}&=& 
                 -k_{\rm m^+ d}(Z)n_{\rm d}(Z) n_{\rm m^+}
                 -k_{\rm M^+ d}(Z) n_{\rm d}(Z) n_{\rm M^+}
                 -k_{\rm e d}(Z) n_{\rm d}(Z) n_{\rm e}\nonumber\\
               &\ \ &+k_{\rm m^+ d}(Z-1) n_{\rm d}(Z-1) n_{\rm m^+}
                 +k_{\rm M^+ d}(Z-1) n_{\rm d}(Z-1) n_{\rm M^+}\nonumber\\
               &\ \ &+k_{\rm e d}(Z+1) n_{\rm d}(Z+1) n_{\rm e},
    \label{224}
\end{eqnarray}
where $n_{\rm j}$ is the number density of each particles  
( $\rm j= m^+: molecular\ ions,\ M^+: heavy\ metal\ ions,\ 
  M: heavy\ metal\ atoms,\ e: electrons,\ 
 d: dust\  grains$), $Z$ is the charge of dust grains, 
$\zeta$ is ionization rate, and $\alpha_{\rm m^+}$, 
$\alpha_{\rm M^+}$, $\beta$, and $k_{\rm j d}(Z)$ are the rate 
coefficients for dissociative recombination, radiative recombination,
charge transfer, and capture of gaseous particles by dust grains, 
respectively.
The timescale of charge transfer is so short that we can neglect 
dissociative recombination and capture of molecular ions by dust grains.
This means that almost all the positive charges are transferred 
to the heavy metal ions. Ignoring the time-derivative term and the 
second and fourth terms on the right hand side of 
Equation $(\ref{221})$, we obtain the relation 
$\beta n_{\rm m^+} n_{\rm M}\simeq\zeta n_{\rm n}$.
Then, we can write the equations simply as follows:
\begin{equation}
    \frac{dn_{\rm M^+}}{dt} = \zeta n_{\rm n} - \alpha_{\rm M^+}n_{\rm M^+}n_{\rm e} 
       -\sum_{Z}^{}k_{\rm M^+ d}(Z)n_{\rm d}(Z) n_{\rm M^+},
    \label{225}
\end{equation}
\begin{equation}
    \frac{dn_{\rm e}}{dt} = \zeta n_{\rm n} - \alpha_{\rm M^+}n_{\rm M^+}n_{\rm e} 
       -\sum_{Z}^{}k_{\rm e d }(Z)n_{\rm d}(Z) n_{\rm e},
    \label{226}
\end{equation}
\begin{eqnarray}
    \frac{dn_{\rm d}(Z)}{dt}&=& 
                 -k_{\rm M^+ d}(Z) n_{\rm d}(Z) n_{\rm M^+}
                 -k_{\rm e d}(Z) n_{\rm d}(Z) n_{\rm e}\nonumber\\
               &\ \ &+k_{\rm M^+ d}(Z-1) n_{\rm d}(Z-1) n_{\rm M^+}
                 +k_{\rm e d}(Z+1) n_{\rm d}(Z+1) n_{\rm e},
    \label{dust}
\end{eqnarray}
In this way, we can treat the reaction equations 
as if $\rm M^+$ are formed directly by primary ionization as in Figure 2.

Numerical calculation of Equation $(\ref{225})-(\ref{dust})$ is not an easy task.
First, as the maximum value of $|Z|$ becomes greater, the number of terms and 
equations increases, and the system of equations become more complicated. 
Second, since the timescales in the system are very different,
it is difficult to solve the equations explicitly in time.
We describe our method to speed up the calculation of these equations
in Sections 2.4 and 2.5.

\subsection{Rate Coefficients}
We use the value of rate coefficients of recombination $\alpha$
and charge transfer $\beta$ 
given by UMIST database (RATE`06), and summarize them in Table1. 
\begin{table}[t]
\centering
\begin{tabular}{ll}
    \hline
    \hline
           Reaction & Rate Coefficient [$\rm cm^3/s$]\\
    \hline
           $ \rm Mg^+ + e^- \ \longrightarrow Mg + h\nu$
            &
           $ \alpha_{\rm M^+}=2.80\times 10^{-12}(T/300)^{-0.86}$\\
           $ \rm   HCO^+ + Mg \ \longrightarrow  \ Mg^+ + HCO$
            &
           $ \beta =2.90\times10^{-9}$ \\ 
    \hline
\end{tabular}
\caption{Rate coefficients given by UMIST database (RATE`06).
$T$ shows the temperature.}
\label{tab:1}
\end{table}
Since we assume that the polarization effect of dust grains is negligible, 
the rate coefficients of gaseous particle capture by dust 
grains are written as
\begin{equation}
    k_{\rm j d}(Z) \equiv \langle \sigma_{\rm j d}(Z) v_{\rm j} \rangle_v,
\label{241}
\end{equation}
where $\langle \ \rangle_v$ means the value averaged by the Maxwellian 
distribution, and $\sigma_{\rm jd}$ is collisional cross section 
between dust grains and gaseous particles: 
\begin{eqnarray}
    \sigma_{\rm j d} =
      \Bigg\{   
   \begin{array}{l}
         S_{\rm j}\pi a^2\left( 1 - \frac{2Q_{\rm j}Q_{\rm d}}
         {am_{\rm j}v^2}\right) \hspace{3.5em} 
         \left( \frac{1}{2}m_{\rm j}v^2 >
                 \frac{Q_{\rm j}Q_{\rm d}}{a}\right)\\
         0 \hspace{10em} 
         \left( \frac{1}{2}m_{\rm j}v^2 <
         \frac{Q_{\rm j}Q_{\rm d}}{a}\right)\\
 \end{array}
\label{242}
\end{eqnarray}
where $Q_{\rm j}$ is the charge of ion or electron, 
$Q_{\rm d}$ is that of dust grains, and $S_{\rm j}$ is the sticking 
probability. We assume $S_{\rm j}=1$ in this paper.
With Equation $(\ref{242})$,
Equation $(\ref{241})$ can be calculated as follows. 
\begin{itemize}
        \item $j=i$ \ (positive ion)
\begin{eqnarray}
    k_{\rm i d}(z) = \Bigg\{  
   \begin{array}{l}
         \pi a^2 \langle v_{\rm i} \rangle_v
         {\rm exp}\left(  {-\frac{q^2 Z}{ak_{\rm B}T}}\right)
         \hspace{2.5em}  (Z>0)\\
         \pi a^2 \langle v_{\rm i} \rangle_v
          \left( 1 - \frac{q^2 Z}{ak_{\rm B}T} \right)
         \hspace{3.2em}  (Z<0)
  \end{array} 
\label{243}
\end{eqnarray}
        \item $j=e$ \ (electron) 
\begin{eqnarray}
    k_{\rm e d}(z) = \Bigg\{
   \begin{array}{l}
         \pi a^2 \langle v_{\rm e} \rangle_v
         \left( 1 + \frac{q^2 Z}{ak_{\rm B}T} \right)\hspace{3.2em} (Z>0)\\
         \pi a^2 \langle v_{\rm e} \rangle_v
         {\rm exp}\left({\frac{q^2 Z}{ak_{\rm B}T}}\right)
            \hspace{3.3em} (Z<0)
    \end{array} 
\label{244}
\end{eqnarray}
\end{itemize}
where $q$ is the charge of an electron, 
$a$ is the radius of the grains, $k_{\rm B}$ is the Boltzmann constant,
and $\langle v_{\rm j} \rangle_v$ is the thermal velocity of the particle: 
\begin{eqnarray}
    \langle v_{\rm j} \rangle_v = \sqrt{\frac{8k_{\rm B}T}{\pi m_{\rm j}}}\ .
\label{245}
\end{eqnarray}
As we can see, particle species dependence is only inverse of the square root of mass.
Thus, we do not have to pay much attention to specific species as long as 
we are interested in ionization degree in a dusty environment.  

In this study, we consider the compact dust grains with the density 
$\rho_{\rm grain}=3\ \rm g\ cm^{-3}$. 
The mass of a dust grains is $m_{\rm grain} = (4\pi/3)\rho_{\rm grain}a^3$,
and the number density of dust grains is $n_{\rm d}= \rho_{\rm d}/
m_{\rm grain}=f_{\rm dg}\rho_{\rm n}/m_{\rm grain}$, where $f_{\rm dg}$ 
is the dust-to-gas mass ratio:
\begin{eqnarray}
    f_{\rm dg} = \frac{\rho_{\rm d}}{\rho_{\rm n}}
\label{247}
\end{eqnarray}
We take $a$ and $f_{\rm dg}$ as free parameters.

\subsection{Gaussian Approximation for Charge Distribution of Dust Grains}
In the basic equations, the third terms of Equation $(\ref{225})$ 
and $(\ref{226})$ contain a number of $Z$-dependent terms. 
This means that the wider the charge distribution we consider, 
the more reactions we have to solve.
Additionally, Equation $(\ref{dust})$ is consist of many equations.
Though we should consider a wide charge distribution for reality, 
it takes a long time to deal with such a large amount of calculation.

\citet{oku09} has shown that charge distribution of dust grains 
can be approximated by a Gaussian distribution.
We confirmed this by solving the basic equations with 
$-4\hspace{0.3em}\raisebox{0.4ex}{$<$}\hspace{-0.75em}
\raisebox{-.7ex}{=}\hspace{0.3em} Z  \hspace{0.3em}
\raisebox{0.4ex}{$<$}\hspace{-0.75em}
\raisebox{-.7ex}{=}\hspace{0.3em}$4.
We use $\zeta=7.6\times10^{-19}{\rm s}^{-1},\ T=280{\rm K},\ 
a=0.1{\rm \mu m},\ {\rm and}\ f_{\rm dg}=10^{-2}$.
Initial condition is $n_{\rm M^+}=0,\ n_{\rm e}=0, 
\ n_{\rm d}(0)=1.11\times10^{3}\ {\rm cm}^{-3},\  
{\rm and}\ n_{\rm d}(Z\neq0)=0$. 
Figure 3 is the plots of the number density of dust grains 
as a function of charge $Z$ and their fit by the Gaussian distribution. 
\begin{figure}[t]
    \epsscale{0.32}
    \plotone{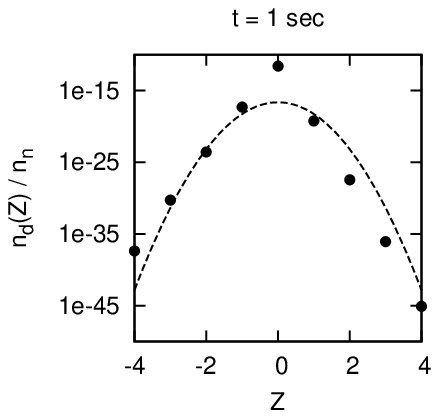}
    \plotone{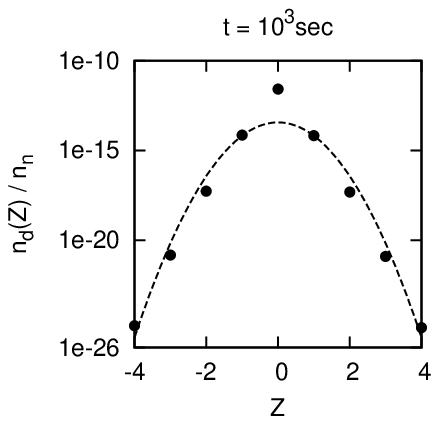}
    \plotone{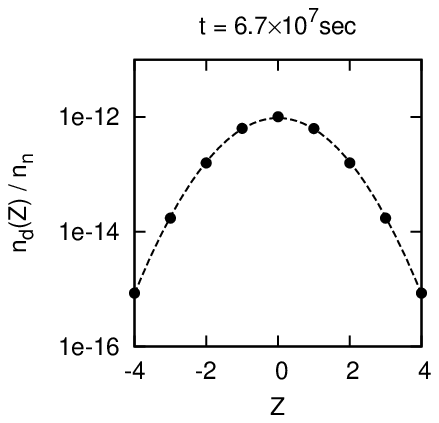}
    \caption{Charge distribution of dust grains. Horizontal axes 
        are charge of dust grains, and vertical axes are the number
        density of each charge of dust grains normalized by the number
        density of neutral gas. Dashed lines are fit by 
        the Gaussian distribution. Initial conditions have the dust grains 
        with only $Z=0$, but the charge distribution already comes 
        close to a Gaussian within only 1 s.  
        At the time ionization degree becomes equilibrium 
        ($t\sim10^7\ \rm s$), charge distribution turns out to be 
        a Gaussian. The middle panel is on the way to be equilibrium.} 
\end{figure}
We can see the charge distribution evolves toward a Gaussian distribution.
Ionization degree reaches equilibrium by $t\sim10^7$ s, and at that time 
the distribution can be well approximated by a Gaussian. 
On the left panel, dust grains tend to be 
charged negatively, because electron can be captured by dust grain 
more quickly than ions; electrons have larger velocity than ions
because they are lighter.  

We define the total number density of dust grains as
\begin{eqnarray}
    N_{\rm d} \equiv \sum_{Z}^{}n_{\rm d}(Z).
\label{250}
\end{eqnarray}
Since it turns out that charge distribution of dust grains can be 
approximated by a Gaussian, $n_{\rm d}(Z)$ can be written as
\begin{eqnarray}
    n_{\rm d}(Z) = \frac{N_{\rm d}}{\sqrt{\mathstrut 2\pi 
                     \langle \delta Z^2 \rangle}}
                     \exp\left[-\frac{(Z - \langle Z \rangle)^{2}}
                     {2\langle \delta Z^2 \rangle}\right]
\label{251}
\end{eqnarray}
The number density of charged dust grains should be obtained 
by calculating the time evolution of the following mean charge of dust grains 
$\langle Z \rangle$ and dispersion of the distribution $\langle \delta Z^2\rangle$:
\begin{eqnarray}
    \langle Z \rangle
    \equiv \frac{1}{N_{\rm d}}\displaystyle\sum_{Z}^{}Zn_{\rm d}(Z)
    \simeq \frac{1}{N_{\rm d}}
        \displaystyle\int_{-\infty}^{\infty}Zn_{\rm d}(Z) ,
\label{252}
\end{eqnarray}
\begin{eqnarray}
    \langle \delta Z^2\rangle
    &\equiv& \langle \left( Z - \langle Z \rangle \right)^2 
                \rangle\nonumber\\
    &=& \langle Z^2 \rangle - \langle Z \rangle ^2 \nonumber\\
    &=& \frac{1}{N_{\rm d}}\displaystyle\sum_{Z}^{}Z^2 n_{\rm d}(Z)
               - \langle Z \rangle ^2 \nonumber\\
    &\simeq& \frac{1}{N_{\rm d}}
            \displaystyle\int_{-\infty}^{\infty}Z^2 n_{\rm d}(Z)
               - \left[ \frac{1}{N_{\rm d}}
                    \displaystyle\int_{-\infty}^{\infty}
                Zn_{\rm d}(Z) \right]^2,
\label{253}
\end{eqnarray}
where $\langle\ \rangle$ means the averaged value weighted by 
the number density of dust grains $n_{\rm d}(Z)$.

First, we take the first moment of Equation $(\ref{dust})$:
\begin{eqnarray}
     \sum_{Z}^{}Z\frac{d}{dt}n_{\rm d}(Z)
        &=& \sum_{Z}^{}\left[ k_{\rm M^+ d}(Z) n_{\rm M^+} n_{\rm d}(Z) 
           - k_{\rm e d}(Z) n_{\rm e} n_{\rm d}(Z) \right]\nonumber\\
        &=& \left[\sum_{Z} k_{\rm M^+ d}(Z) n_{\rm d}(Z)\right]
              n_{\rm M^+}
           - \left[\sum_{Z} k_{\rm e d}(Z) 
               n_{\rm d}(Z)\right] n_{\rm e} \nonumber\\
        &=& \langle k_{\rm M^+ d}\rangle N_{\rm d} n_{\rm M^+} 
           -\langle k_{\rm e d} \rangle N_{\rm d} n_{\rm e},
    \label{254}
\end{eqnarray}
where $\langle k_{\rm M^+d} \rangle$ and $\langle k_{\rm ed} \rangle$ 
are averaged rate coefficients weighted by $n_{\rm d}(Z)$, and
their derivation is shown in the Appendix.
In \citet{oku09}, the positively charged cases of $k_{\rm jd}(Z)$ are 
ignored because dust grains tend to be charged negatively as they grow. 
In this study, positively charged cases are also considered.
This is because the inclusion of positively charged cases slightly
affects the results.
With Equation ($\ref{254}$) and 
\begin{eqnarray}
    \frac{d}{dt}\langle Z \rangle 
        &=& \frac{d}{dt}\left[\frac{1}{N_{\rm d}}
                 \displaystyle\sum_{Z}Zn_{\rm d}(Z)
                 \right]\nonumber\\
        &=& \frac{1}{N_{\rm d}}\left[\sum_{Z}^{}{Z}
                \frac{d}{dt}n_{\rm d}(Z)\right],
\label{257}
\end{eqnarray}
we can derive the differential equation of $\langle Z \rangle$:
\begin{equation}
    \frac{d\langle Z \rangle}{dt}=
        \langle k_{\rm M^+ d} \rangle n_{\rm M^+} 
        -\langle k_{\rm e d} \rangle n_{\rm e}. 
\label{258}
\end{equation}

Furthermore, we take the second moment of Equation $(\ref{dust})$:
\begin{eqnarray}
     \sum_{Z}^{}Z^2\frac{d}{dt}n_{\rm d}(Z)
        &=& \langle k_{\rm M^+ d} \rangle N_{\rm d} n_{\rm M^+} 
           +2\langle Zk_{\rm M^+ d} \rangle 
           N_{\rm d} n_{\rm M^+}\nonumber\\
          &\ \ & + \langle k_{\rm e d} \rangle N_{\rm d} n_{\rm e}
           -2\langle Zk_{\rm e d} \rangle N_{\rm d} n_{\rm e},
\label{259}
\end{eqnarray}
In a similar way, we obtain the following equation:
\begin{eqnarray}
    \frac{d}{dt}\langle Z^2 \rangle
        &=& \langle k_{\rm M^+ d} \rangle n_{\rm M^+} 
           +2\langle Z k_{\rm M^+ d} \rangle n_{\rm M^+}\nonumber\\ 
          &\ \ & + \langle k_{\rm e d} \rangle n_{\rm e} 
           -2\langle Z k_{\rm e d} \rangle n_{\rm e} .
    \label{2512}
\end{eqnarray}
Finally, we can derive the differential equation of 
$\langle \delta Z^2\rangle$:
\begin{eqnarray}
    \frac{d\langle \delta Z^2 \rangle}{dt}
   &=& \frac{d\langle Z^2 \rangle}{dt} 
        - \frac{d\langle Z \rangle^2}{dt}\nonumber\\
   &=& \frac{d\langle Z^2 \rangle}{dt} - 2\langle Z \rangle 
         \frac{d\langle Z \rangle}{dt}\nonumber\\
   &=& \left( \langle k_{\rm M^+ d} \rangle 
        +2\langle Z k_{\rm M^+ d} \rangle
        -2\langle Z \rangle \langle k_{\rm M^+ d} \rangle
        \right)n_{\rm M^+}\nonumber\\
   &\ \ & + \left( \langle k_{\rm e d} \rangle 
        -2\langle Z k_{\rm e d} \rangle
        +2\langle Z \rangle\langle k_{\rm e d} \rangle \right)
        n_{\rm e}\nonumber\\
   &=& \left( \langle k_{\rm M^+ d} \rangle 
           +2\langle k_{\rm M^+ d} \delta Z \rangle \right)n_{\rm M^+}
      +\left( \langle k_{\rm e d} \rangle
            -2\langle k_{\rm e d} \delta Z \rangle \right)n_{\rm e}.
        \label{2513}
\end{eqnarray}
The formulae of $\langle k_{\rm M^+d} \delta Z \rangle$ and 
$\langle k_{\rm M^+d} \delta Z \rangle$ are written in Appendix.
By this Gaussian approximation, the number of equations can be 
dramatically reduced.

\subsection{Acceleration by Piecewise Exact Solution}
The equations we have to solve are as follows:
\begin{eqnarray}
    \frac{dn_{\rm M^+}}{dt} &=& \zeta n_{\rm n} - \alpha_{\rm M^+}n_{\rm M^+}n_{\rm e} 
       -\langle k_{\rm M^+ d}\rangle N_{\rm d} n_{\rm M^+},\\
    \label{261}
    \frac{dn_{\rm e}}{dt} &=& \zeta n_{\rm n} - \alpha_{\rm M^+}n_{\rm M^+}n_{\rm e} 
       -\langle k_{\rm e d }\rangle N_{\rm d} n_{\rm e},\\
    \label{262}
    \frac{d\langle Z \rangle}{dt}&=&
        \langle k_{\rm M^+ d} \rangle n_{\rm M^+} 
        -\langle k_{\rm e d} \rangle n_{\rm e}, \\
\label{263}
%
    \frac{d\langle \delta Z^2 \rangle}{dt}
    &=& \left( \langle k_{\rm M^+ d} \rangle 
           +2\langle  k_{\rm M^+ d} \delta Z \rangle \right)n_{\rm M^+}\nonumber\\
    &\ \ &  +\left( \langle k_{\rm e d} \rangle
            -2\langle k_{\rm e d} \delta Z \rangle \right)n_{\rm e}.
\label{264}
\end{eqnarray}
We solve our equations 
partially with piecewise exact solution \citep{ino08}.
This method enables us to solve with large time steps.

First, we split our equations and solve analytically. 
The solution of 
\begin{eqnarray}
    \frac{dn_{\rm M^+}}{dt} 
        = \zeta n_{\rm n} -\langle k_{\rm M^+ d}\rangle N_{\rm d} n_{\rm M^+},
\label{265}
\end{eqnarray}
is 
\begin{eqnarray}
   n_{\rm M^+}(t+\Delta t) 
       &=& \left( n_{\rm M^+}(t)
             - \frac{\zeta n_{\rm n}}{\langle k_{\rm M^+ d}\rangle N_d}\right) 
        e^{-\langle k_{\rm M^+ d}\rangle N_d \Delta t} 
        + \frac{\zeta n_{\rm n}}{\langle k_{\rm M^+ d}\rangle N_d},
 \label{266}
\end{eqnarray}
where $\Delta t$ is the time step of time integration, 
and the solution of
\begin{equation}
    \frac{dn_{\rm e}}{dt}
        = \zeta n_{\rm n} -\langle k_{\rm e d }\rangle N_{\rm d} n_{\rm e}
 \label{267}
\end{equation}
is 
\begin{eqnarray}
    n_{\rm e}(t+\Delta t)
        &=& \left( n_{\rm e}(t) - \frac{\zeta n_{\rm n}}
                {\langle k_{\rm e d }\rangle N_d}\right)
           e^{-\langle k_{\rm e d }\rangle N_d \Delta t} 
           + \frac{\zeta n_{\rm n}}{\langle k_{\rm e d }\rangle N_d}. 
 \label{268}
\end{eqnarray}
We treat rate coefficients such as $\langle k_{j \rm d}\rangle$
as constants during one step.
From Equation $(\ref{261})-(\ref{263})$,
\begin{eqnarray}
    \frac{dn_{\rm e}}{dt} - \frac{dn_{\rm M^+}}{dt}        
        &=& \langle k_{\rm M^+ d} \rangle n_{\rm M^+}N_{\rm d} 
        -\langle k_{\rm e d} \rangle n_{\rm e}N_{\rm d},\nonumber\\
        &=& N_{\rm d}\frac{d\langle Z \rangle}{dt},
\label{269}
\end{eqnarray}
and with charge conservation, 
\begin{eqnarray}
    N_{\rm d}\langle Z\rangle(t) +  
     n_{\rm M^+}(t) - n_{\rm e}(t)= {\rm constant} = 0, 
\label{2610}
\end{eqnarray}
\begin{eqnarray}
    \langle Z\rangle(t+\Delta t) = \langle Z\rangle(t) 
    + \frac{n_{\rm e}(t + \Delta t) - n_{\rm M^+}(t + \Delta t)}{N_{\rm d}} 
    - \frac{n_{\rm e}(t) - n_{\rm M^+}(t)}{N_{\rm d}}. 
\label{2611}
\end{eqnarray}
The solution of
\begin{eqnarray}
    \left\{
        \begin{array}{l}
            \displaystyle\frac{dn_{\rm M^+}}{dt} 
             = - \alpha_{\rm M^+}n_{\rm M^+}n_{\rm e}\\ 
            \displaystyle\frac{dn_{\rm e}}{dt} 
             = - \alpha_{\rm M^+}n_{\rm M^+}n_{\rm e} 
        \end{array}
    \right.
    \label{2612}
\end{eqnarray}
with Equation ($\ref{2610}$) is 
\begin{eqnarray}
    \left\{
        \begin{array}{l}
    n_{\rm M^+}(t+\Delta t)=
        \displaystyle\frac{N_{\rm d}\langle Z\rangle(t)}
           {(N_{\rm d}\langle Z\rangle(t)/n_{\rm M^+}(t) + 1)
               \exp[\alpha_{\rm M^+}N_{\rm d}\langle Z\rangle(t)\Delta t]-1}\\
    n_{\rm e}(t+\Delta t)=
        n_{\rm M^+}(t+\Delta t) + N_{\rm d}\langle Z\rangle(t).
        \end{array}
    \right.
\label{2613}
\end{eqnarray}
We use these solutions as initial conditions of the time integration of
the remaining complex equation:
\begin{eqnarray}
    \frac{d\langle \delta Z^2 \rangle}{dt}
    &=& \left( \langle k_{\rm M^+ d} \rangle 
          +2\langle k_{\rm M^+ d} \delta Z \rangle \right)n_{\rm M^+}
          \nonumber\\
    &\ \ & +\left( \langle k_{\rm e d} \rangle
            -2\langle k_{\rm e d} \delta Z \rangle \right)n_{\rm e}.
\label{2614}
\end{eqnarray}

In conclusion, what to be solved numerically are Equations $(\ref{266})$,
$({\ref{268}})$, $(\ref{2611})$, $(\ref{2613})$, and $(\ref{2614})$.

\section{Test Calculation on Protoplanetary Disks}
We compare the result of our fast calculation method with the result 
of the direct calculation of the original equations $(\ref{225}),\ 
(\ref{226}),\ {\rm and}\ (\ref{dust})$ in order to find out how accurate our method is.

\subsection{Disk Property of Protoplanetary Disks}
  We adopt the disk property of the minimum mass solar nebula 
(MMSN) model \citep{hay81}.
The surface density and the temperature of the gas in the disk are  
\begin{eqnarray}
    \Sigma_{\rm n} = 1.7\times 10^3
    \left( \frac{r}{{\rm 1AU}} \right)^{-1.5}\ {\rm g\ cm^{-2}},
\label{31}
\end{eqnarray}
\begin{eqnarray}
    T = 280\left( \frac{r}{{\rm 1AU}} \right)^{-0.5}\ {\rm K},
\label{32}
\end{eqnarray}
where $r$ is the orbital radius.
The density of the disk is defined as follows:
\begin{eqnarray}
    \rho_{\rm n}(r,z) \equiv \frac{\Sigma_{\rm g}}
        {\sqrt{\mathstrut 2\pi}H}
        \exp\left( -\frac{z^2}{2 H^2} \right),
\label{33}
\end{eqnarray}
where $H$ is the scale height of the disk:
\begin{eqnarray}
     H \equiv \frac{c_{\rm s}}{\Omega_{\rm k}},
    \label{}
\end{eqnarray}
where $c_{\rm s} = \sqrt{\mathstrut \gamma k_{\rm B}T/\mu m_{\rm H}}$ is 
the sound speed, $\Omega_{\rm k}=
\sqrt{\mathstrut GM_\ast/r^3}$ is the Keplerian
frequency. 
Here, $\gamma$ is the specific heat ratio, $\mu$ is the mean molecular weight 
of the neutral gas, $m_{\rm H}$ is the mass of a hydrogen atom, 
$G$ is the gravitational constant, and $M_\ast$ is the mass of the central star.
We adopt $f_{\rm dg}=10^{-2}$, $\mu=2.34$, and $M_\ast=M_\odot$. 

\subsection{Ionization Rate}
There are various primary ionization sources, for example,  
Galactic cosmic rays, UV and X-rays from central stars, decay of 
radionuclide, and thermal ionization, etc.
Here, just simplicity, we neglect the possible contribution of energetic 
electrons \citep{inu05}, and only consider cosmic rays, X-rays, and radionuclide. 
With orbital radius $r$ and perpendicular oriented length $z$, 
ionization rate is written as follows:
\begin{eqnarray}
    \zeta(r,z) &\simeq& \zeta_{\rm C} + \zeta_{\rm X}
                    + \zeta_{\rm R},
\label{231}
\end{eqnarray}
where $\zeta_{\rm C}$, $\zeta_{\rm X}$, {\rm and}\ $\zeta_{\rm R}$ are the ionization 
rate by cosmic rays, stellar X-rays, and radionuclide, respectively.
$\zeta_{\rm C}$ is given by the following equation:
\begin{eqnarray}
    \zeta_{\rm C} &\simeq& \frac{\zeta_{\rm CR}}{2}
    \left\{ 
        \exp \left[-\frac{\chi(r,z)}{\chi_{\rm CR}}\right]
    \right.
         \left.+ \exp\left[ - \frac{\Sigma(r) 
            - \chi(r,z)}{\chi_{\rm CR}} \right]     
    \right\},
\label{232}
\end{eqnarray}
where $\zeta_{\rm CR}\sim1.0\times10^{-17}\ {\rm s^{-1}}$ is 
the ionization rate by cosmic rays in the interstellar space,
and $\chi_{\rm CR}\sim96\ {\rm g\ cm^{-2}}$
is the attenuation length of the ionization rate by cosmic rays.
\begin{eqnarray}
    \chi(r,z) = \int_{z}^{\infty}\rho(r,z)dz\ \ \rm [g\ cm^{-2}],
\label{233}
\end{eqnarray}
is the vertical column density of the gas measured from the outside of 
the disk \citep{ume81}.
The ionization rate by X-ray is given as follows:
\begin{equation}
    \zeta_{\rm X} \simeq \zeta_{\rm XR}\left(\frac{r_{\ast}}{1{\rm AU}}\right)^{-2}
                    \left( \frac{L_{\rm XR}}{2\times10^{30}{\rm erg\ s^{-1}}} \right)
                    \left\{ \exp \left[-\frac{\chi(r,z)}{\chi_{\rm XR}}\right]
                    + \exp\left[ -\frac{\Sigma(r) -\chi(r,z)}{\chi_{\rm XR}}\right]
                    \right\},
    \label{234}
\end{equation}
where $r_{\ast}$ is the distance from the star, 
$L_{\rm XR}\sim 2\times 10^{30}$ erg s$^{-1}$ is the X-ray luminosity, and 
$\zeta_{\rm XR}\sim2.6\times10^{-15}$ s$^{-1}$ and 
$\chi_{\rm XR}\sim 8.0$ g cm$^{-2}$ are the fitting parameters 
(\citealt{ige99}; \citealt{tur08}). 
We adopt ionization rate by radionuclide $\zeta_{\rm R}=7.6\times
10^{-19}\ {\rm s}^{-1}$ \citep{ume09}. 

\subsection{Result of the Test Calculation}
We assume cosmic rays, stellar X-rays, and radionuclide as ionization sources,
and calculate at the mid-plane of the orbital radius 1 AU 
from the central star. The number density of neutral gas is
$n_{\rm n}=\rho_{\rm n}(1\rm AU, 0)/\mu m_{\rm H} =4.22\times10^{14}\ \rm cm^{-3}$, 
and the ionization rate is 
$\zeta = \zeta(1\rm AU,0) \simeq7.6\times10^{-19}\ \rm s^{-1}$.
We assume the disk is isothermal in the vertical direction, and
use $\gamma=1$ in the expression of $c_{\rm s}$. 
\begin{figure}[h]
    \epsscale{1.00}
    \plottwo{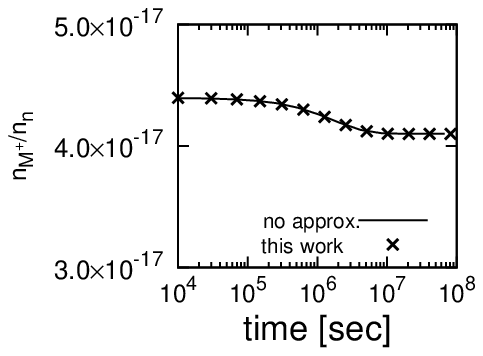}{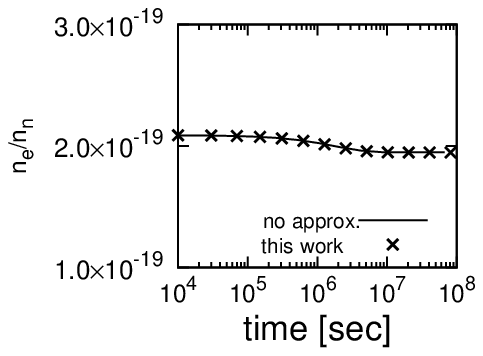}\\
    \epsscale{1.00}
    \plottwo {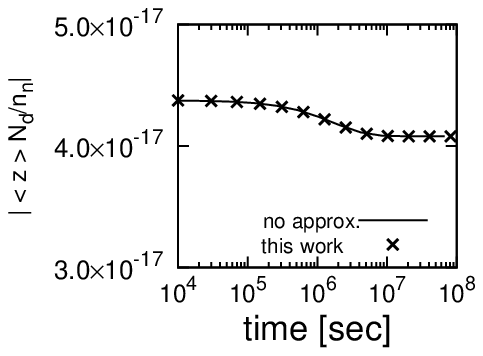}{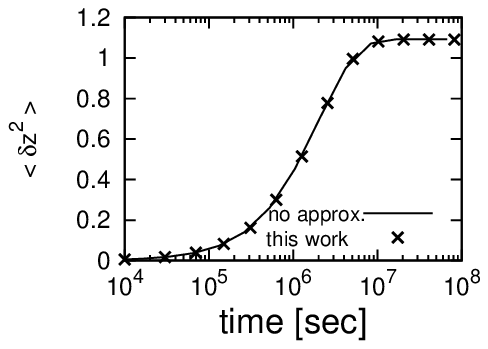}\\
    \caption{Top-left panel is the number density of heavy metal ion,
       top-right panel is that of free electron, and both are 
       normalized by the number density of neutral gas. The bottom-left 
       panel is the absolute value of the product of averaged 
       dust charge and the total abundance of dust grains---this 
       means the total charge which dust grains have; and the bottom-right
       panel is the dispersion of the dust charge distribution.
       Lines show the results of the basic equations, and crosses show 
       the results of our calculation with two kinds of speed-up device. 
       These panels show that the Gaussian distribution approximation and 
       piecewise exact solution are successful.}
    \end{figure}

Figure 4 shows the result of the comparison, and we can see that they 
agree very well.
With our method, we can calculate with about $10^5$ times larger  
time steps. This means that calculation becomes dramatically fast.

We start our calculation with the initial condition 
$x_{\rm M^+}=0, x_{\rm e}=0, {\rm and}\ \langle Z\rangle=0$.
It takes about $10^7$ s to come to be equilibrium.
This is the timescale for charged particles to meet dust grains.
\citet{oku09} has shown that this timescale is written as
\begin{eqnarray}
    t^{-1} \simeq \frac{\zeta n_{\rm n}}{n_{\rm d}}\ \ [{\rm s}^{-1}].
\label{}
\end{eqnarray}
The existence of such long timescale events indicates that 
our time-dependent method is useful.

\section{Ionization Degree in Circumplanetary Disks}
Circum-planetary disks are formed after planet formation and 
thought to be the site of satellite formation (e.g., \citealt{can02} 
; \citealt{sas10}). Understanding of the evolutions of 
circumplanetary disks is important in the context of satellite formation.
One of the important factors in disk evolution is the ionization 
degree which has been calculated by \citet{tak96},
although, the surface density used in their study seems to 
be too heavy, and the effect of dust grains are not concerned.
In this Section, we calculate the ionization degree in circumplanetary 
disks including the charged particle capture by dust grains.

\subsection{Disk Property of Circumplanetary Disks}
We use an actively supplied gaseous accretion disk model
(\citealt{can02}, \citealt{can06}; see Appendix of \citealt{sas10})
The temperature is
\begin{eqnarray}
    T_{\rm cir} \simeq 160\left( \frac{M_{\rm p}}{M_{\rm J}} \right)^{1/2}
   \left( \frac{r}{20R_{\rm J}} \right)^{-3/4}\ \rm K, 
\label{}
\end{eqnarray}
where $M_{\rm p}$ is the mass of the central planet, 
$M_{\rm J}$ is that of Jupiter,  
$r$ is the orbital radius from the central planet, and $R_{\rm J}$ 
is the radius of Jupiter. In their model, we adopt $5\times10^6\ $yr 
as the accretion timescale for gaseous disk as heavy as the central 
planet, and assume the disk is vertically isothermal.  
The surface density is given as follows with viscosity coefficient 
$\alpha=5\times10^{-3}$: 
\begin{equation}
    \Sigma_{\rm cir} \simeq 100\left( \frac{M_{\rm p}}{M_{\rm J}} 
             \right)\left( \frac{r}{20R_{\rm J}} \right)^{-3/4}\ 
    \rm g\ cm^{-2}. 
\label{}
\end{equation}

\subsection{Result}
We apply our method for the calculation of the ionization
degree in the circumplanetary disks.
We assume $M_{\rm p}=M_{\rm J}$.
The abundance of radionuclide is uncertain, but ionization by cosmic 
rays is efficient, and radionuclide dose not 
affect the ionization degree very much. 
Since circumplanetary disks are located at orbital radius of gas 
giant planet which is not so close to the star, 
X-ray ionization is less effective. Furthermore, the scale height of 
circumplanetary disks is far smaller than that of protoplanetary disks. 
So, geometrically, it is difficult for X-rays to reach the 
circumplanetary disks. When we think about circumplanetary disks, 
we consider only cosmic ray ionization.

In this study, we investigate the extent of the MRI-inactive regions in the 
circumplanetary disks.
We parameterize the vertical component of plasma beta (at mid-plane),
the ratio of 
gas pressure and magnetic pressure, $\beta=P_{\rm gas}/P_{\rm mag}$,
the radius of dust grains $a$,
 and the dust-to-gas mass ratio $f_{\rm dg}=\rho_{\rm d}/\rho_{\rm g}$.
Figure 5 shows the size of the MRI-inactive region of the model $\beta=100,\ 
a=10^{-3},\ {\rm and}\ f_{\rm dg}=10^{-2}$. As a matter of convenience for the 
discussion of the MRI, we take $z/H$ as $y$ axes. 
Figures 6 and 7 show the extent of the MRI-inactive regions. We use Elsasser number 
as magnetic Reynolds number $Re_{\rm m}$:
\begin{eqnarray}
   Re_{\rm m}  = \frac{v_{\rm Az}^2}{\eta \Omega_{\rm K}},
\label{411}
\end{eqnarray}
where $v_{\rm Az}$ is the vertical component of Alfv\'en velocity.
If $Re_{\rm m}<1$, MRI dose not happen. 
%
%

\begin{figure}[t]
    \epsscale{0.60}
    \plotone{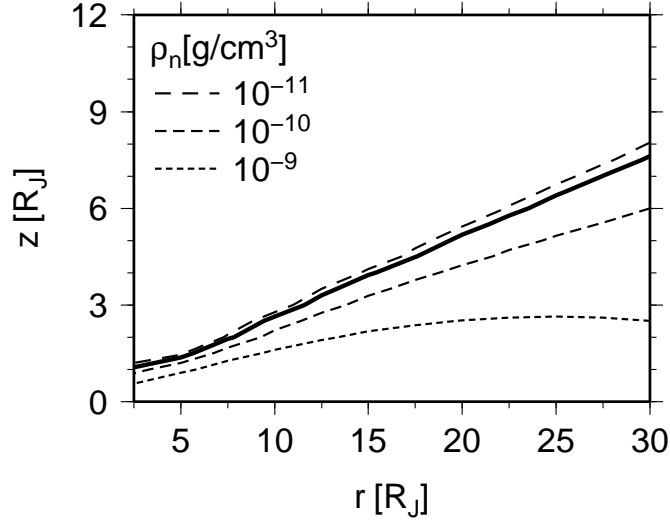}
    \caption{Size of the MRI-inactive region with $\beta=100$, $a=10^{-3}$cm,
        {\rm and}\ $f_{\rm dg}=10^{-2}$. The horizontal axis is orbital radius
        from the central planet and vertical axis is vertical extent
        of the disk. The region under the black line is the MRI-inactive region.
        The density of the neutral gas is also shown as contour lines.}
    \end{figure}

\begin{figure}
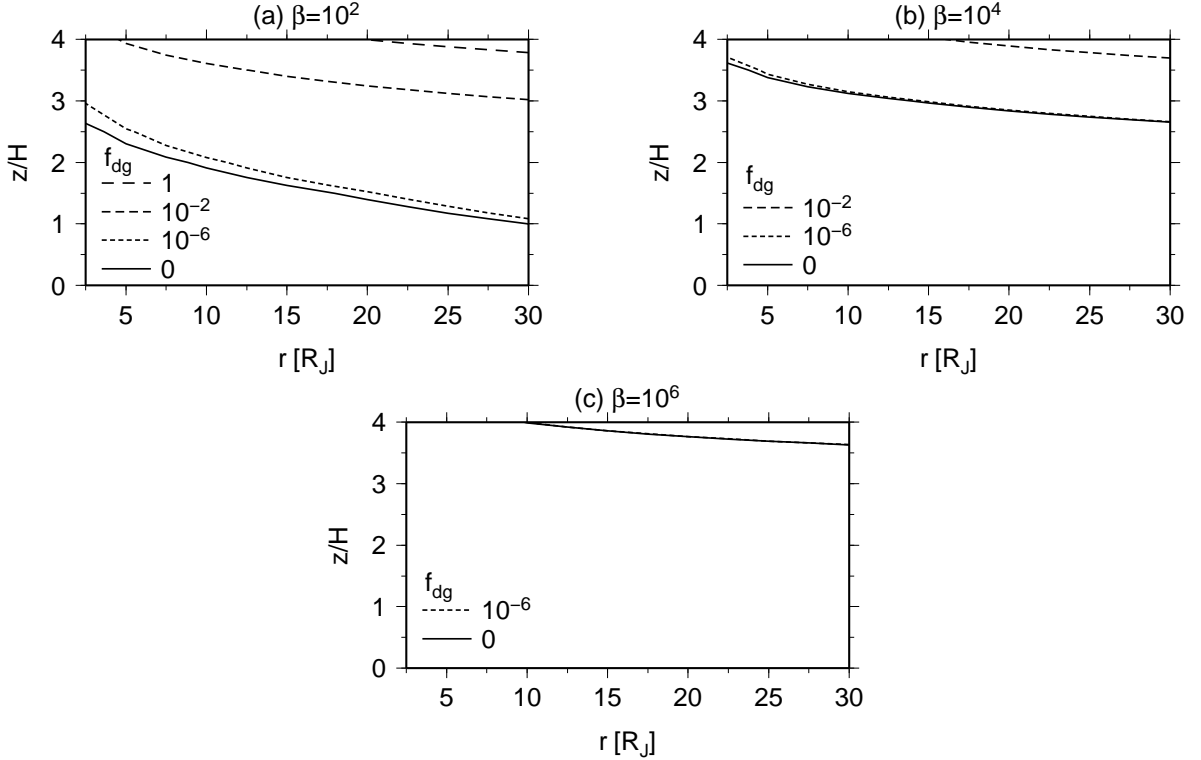

    \epsscale{1.00}
    \plottwo{sate-beta2--a-3.eps}{sate-beta4--a-3.eps}\\
    \epsscale{0.50}
    \plotone{sate-beta6--a-3.eps}
    \caption{Boundaries of the MRI-inactive region in circumplanetary disks: 
     (a)$\beta=10^2$, (b)$\beta=10^4$, {\rm and}\ (c)$\beta=10^4$.
     The horizontal axis denotes the orbital radius from the 
     central planet, and the vertical axis is vertical extent of 
     the disk that is normalized by the scale height of the corresponding 
     radius.  The radius of dust grains is $a=10\mu$m.
     Contour lines show the boundaries of the MRI-active region and the 
     MRI-inactive region
     for each model of the dust-to-gas mass ratio $f_{\rm dg}$. 
     The circumplanetary disks are not expected to be magnetically active.}
    \end{figure}


\begin{figure}
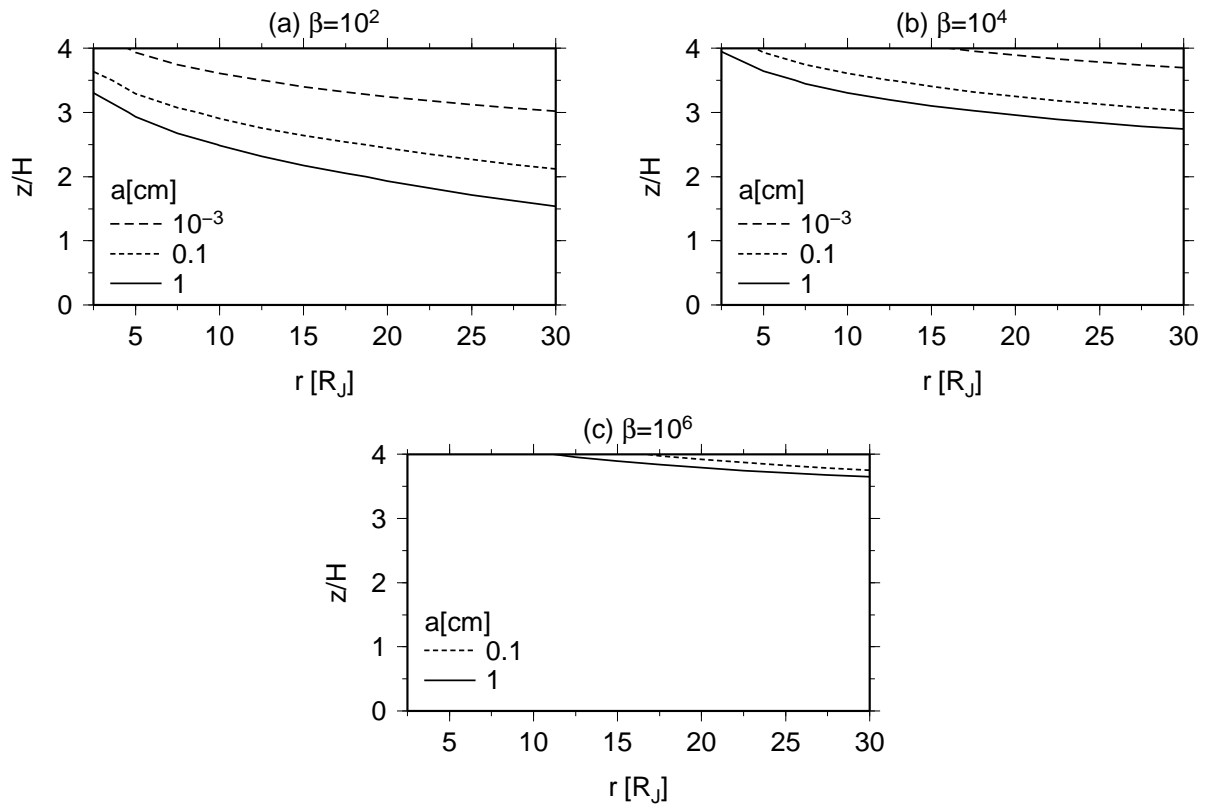

   \begin{center}
   \begin{tabular}{cc}
    \includegraphics[width=8cm]{sate-beta2--R-2.eps}
    &
    \includegraphics[width=8cm]{sate-beta4--R-2.eps}\\
   \end{tabular}\\
    \includegraphics[width=8cm]{sate-beta6--R-2.eps}
   \end{center}
    \caption{Same as Figure 6, but the dust-to-gas mass ratio is 
     fixed to $f_{\rm dg}=10^{-2}$, and consider the dust grain radius $a$
     cm as a parameter.} 
\end{figure}

In Figure 6, we can see the $f_{\rm dg}$ dependence of the size of
the MRI-inactive regions. Figure 7 shows the size of the MRI-inactive regions of various
model of $a$.
We also calculate the models of $f_{\rm dg}=10,\ 100$.
The result of the model $a=$1mm for various value of $f_{\rm dg}$ is
summarized in Table 2.  Our method can calculate even such large mean 
charge and dispersion of dust grains.

\begin{table}[t]
\centering
\begin{tabular}{llll}
    \hline
    \hline
     $f_{\rm dg}$ & $n_{\rm e}/n_{\rm n}$ & $\langle Z \rangle$ 
     & $\langle \delta Z^2\rangle$ \\
    \hline
     100 & 4.83$\times10^{-19}$ & $-2.26\times10^3$ & 6.46$\times10^3$ \\
     10 & 8.97$\times10^{-17}$ & $-3.70\times10^4$ & 9.56$\times10^3$ \\ 
     1 & 1.75$\times10^{-15}$ & $-4.49\times10^4$ & 9.83$\times10^3$ \\
     $10^{-2}$ & 1.76$\times10^{-13}$ & $-4.50\times10^4$ & 9.83$\times10^3$ \\ 

    \hline
    \hline
\end{tabular}
\caption{Ionization degree $n_{\rm e}/n_{\rm n}$, mean charge of dust grains 
    (normalized by the charge of an electron) $\langle Z \rangle$, 
    and dispersion of the charge distribution of dust grains 
     $\langle \delta Z^2\rangle$ 
    at $r=15R_{\rm J},\ z=0$. The radius of dust grains is fixed at $a=1$mm 
    and take dust to gas mass ratio $f_{\rm dg}$ as a parameter. Our method 
    is effective even if the charge distribution of dust grains is very wide.}
\label{tab:2}
\end{table}

Our result indicates that almost entire parts of 
circumplanetary disks are MRI-inactive.

\subsection{Discussion}
Previous investigation on circumplanetary disks is mainly focused 
on active disks for theoretical reasons, but all regions in 
circumplanetary disks we investigated turn out to be the MRI-inactive region. 
This suggests that MRI does not occur for a long time, and satellites 
can be formed slowly, perhaps by gravitational collapse.
The temperature of the disk might be lower because viscous heating
is not so efficient.
However, there are many other accretion mechanisms such as gravitational
torques and spiral waves \citep[e.g.,][]{mac10}. We would like to consider these effects in the
future work.

There remains various uncertainty on circumplanetary disks,
for example, how many dust grains remain after planet formation,
and how large they are. We have to study ionization degree in various 
cases.
We also have to investigate it using other disk models.
In this work, we suppose only fixed size of dust grains, so 
the result may change if we consider the coagulation of dust grains;
or if we consider ice as dust grains, result also may change. 

\section{Summary}
We have developed a fast and accurate method to calculate the ionization 
degree in protoplanetary and circumplanetary disks.
This method can calculate it time dependently by using the following 
two kinds of speed-up device.\\
1)Gaussian approximation for the charge distribution of dust grains.
We have confirmed that the charge distribution of dust grains can be 
approximated by Gaussian distribution \citep{oku09} by solving 
basic reaction equations. This approximation can reduce the number 
of equations.\\
2)Piecewise exact solution.
We have used piecewise exact solution that is developed by \citet{ino08}.
We solve only rapid reaction terms analytically in 
advance, then we solve remaining terms numerically using the analytic
solutions as an initial condition of time integration.
This method enables us to calculate without limitation of time step
by rapid reactions. 
Since our method can calculate the ionization degree accurately and
very rapidly, we would like to plug our method into MHD simulations.

We have checked our calculation method by comparing the result 
with the direct calculation of basic reaction equations and 
shown that they agree very well.

We have applied our calculation method for circumplanetary disks.
Our method can calculate ionization degree quickly even when the 
dispersion of dust grains is about $10^{4}$.
We have investigated with model parameters $a=10^{-5},\ 10^{-3},\  
0.1,\ 1$ cm, $f_{\rm dg}=10^{-6},\ 10^{-2},\ 1,\ 10,\ 100$, and
$\beta=10^{2},\ 10^{4},\ 10^{6}$.
The results show that almost all regions of circumplanetary disks
are the MRI-inactive regions.
This suggests that gas in circumplanetary disks accrete more slowly 
than previously thought.

\acknowledgments

We would like to thank the referee for his/her useful comments.
We also thank Dr. Takeru K. Suzuki and Dr. Kazunari Iwasaki for their 
constructive advice and helpful discussion.
S.O. is supported by a Grant-in-Aid for JSPS Fellows 
($22\cdot 7006$) from the MEXT of Japan, and S.I. is 
supported by Grant-in-Aid for Scientific Research from the 
MEXT of Japan (23244027).

\appendix

\section{Derivation of Rate Coefficients}
From Equation $(\ref{243})$, the rate coefficient of the metal ion 
capture by dust grains averaged by charge of dust grains is defined as 
follows:
\begin{eqnarray}
    \langle k_{\rm M^+d} \rangle
       &\equiv& \frac{1}{N_{\rm d}}\sum_{Z}k_{\rm M^+d}(Z)n_{\rm d}(Z)\nonumber\\
       &\simeq& \frac{1}{N_d}\pi a^2 \langle v_{\rm M^+} \rangle_v
        \left\{
          \int_{-\infty}^{0}\left( 1-\frac{q^2 Z}{a k_{\rm B}T} \right)
          \frac{N_{\rm d}}{\sqrt{\mathstrut 2\pi 
                \langle \delta Z^2 \rangle}}
          \exp\left[-\frac{(Z - \langle Z \rangle)^{2}}
               {2\langle \delta Z^2 \rangle}\right]dZ \right. \nonumber\\
        &\ \ &\left. +\int_{0}^{\infty}
          \exp\left[-\frac{q^2 Z}{a k_{\rm B}T}\right]
          \frac{N_{\rm d}}{\sqrt{\mathstrut 2\pi 
                             \langle \delta Z^2 \rangle}}
          \exp\left[-\frac{(Z - \langle Z \rangle)^{2}}
                      {2\langle \delta Z^2 \rangle}\right]dZ
          \right\} \nonumber\\
        &=&\pi a^2  \langle v_{\rm M^+} \rangle_v
         \left\{
           \frac{1}{2}\left( 1 - \frac{q^2}{a k_{\rm B}T}
                                  \langle Z \rangle \right)
           {\rm erfc}\left[ \frac{\langle Z \rangle}
                 {\sqrt{\mathstrut 2\langle \delta Z^2 \rangle}}
                               \right]\right. \nonumber\\
        &\ \ & + \frac{q^2}{a k_{\rm B}T}
           \sqrt{\mathstrut \frac{\langle \delta Z^2 \rangle}{2\pi}}
           \exp\left[ -\frac{\langle Z \rangle^{2}}
                            {2\langle \delta Z^2 \rangle}\right]\nonumber\\
        &\ \ & \left.+ \frac{1}{2}{\rm erfc} 
                \left[
                  \sqrt{\mathstrut \frac{\langle \delta Z^2 \rangle}{2}}
                     \frac{q^2}{a k_{\rm B}T} 
                 - \frac{\langle Z \rangle}
                       {\sqrt{\mathstrut 2\langle \delta Z^2 \rangle}}
                \right]
            \exp \left[ \frac{1}{2}
                        \left( \frac{q^2} {a k_{\rm B}T} \right)^2
                           \langle \delta Z^2 \rangle
                      - \left( \frac{q^2}{a k_{\rm B}T} \right)
                           \langle Z \rangle  \right]
        \right\},
    \label{255}
\end{eqnarray}
In the same way, we can gain the rate coefficient of electron capture 
averaged by $Z$ from Equation$(\ref{244})$ as the following: 
\begin{equation}
    \langle k_{\rm e d} \rangle 
    \simeq \langle k_{\rm M^+ d} \rangle 
        \mid _{\langle Z \rangle\  \rightarrow\  -\langle Z \rangle} 
    \label{256}
\end{equation}

We show only the result of derivation of 
$\langle k_{\rm M^+d} \delta Z \rangle$ and 
$\langle k_{\rm ed} \delta Z \rangle$ as follows:
\begin{eqnarray}
    \langle k_{\rm M^+d} \delta Z \rangle
    &\simeq& \pi a^2 \langle v_{\rm M^+} \rangle_v
       \frac{q^2}{2ak_{\rm B}T}\langle \delta Z^2 \rangle
       \left(- {\rm erfc}\left[\frac{\langle  Z \rangle}
               {\sqrt{\mathstrut 2\langle \delta Z^2 \rangle}} \right]
       \right. \nonumber\\
       &\ \ &  - \exp \left[- \frac{q^2}{ak_{\rm B}T}\langle Z \rangle 
              + \frac{1}{2}\left(\frac{q^2}{ak_{\rm B}T}\right)^2
              \langle \delta Z^2 \rangle \right]
              \nonumber\\
       &\ \ &  \times {\rm erfc}\left[\frac{q^2}{ak_{\rm B}T}
         \sqrt{\mathstrut\frac{\langle \delta Z^2 \rangle}{2}}
               \left.  - \frac{\langle Z \rangle}
                 {\sqrt{\mathstrut 2\langle \delta Z^2 \rangle}} \right]
      \right),
\label{2510}
\end{eqnarray}
\begin{eqnarray}
    \langle k_{\rm ed} \delta Z\rangle
    &\simeq& \pi a^2 \langle v_{\rm e} \rangle_v
       \frac{q^2}{2ak_{\rm B}T}\langle \delta Z^2 \rangle
       \left({\rm erfc}\left[-\frac{\langle  Z \rangle}
               {\sqrt{\mathstrut 2\langle \delta Z^2 \rangle}} \right]
       \right. \nonumber\\
       &\ \ &  + \exp \left[\frac{q^2}{ak_{\rm B}T}\langle Z \rangle 
              + \frac{1}{2}\left(\frac{q^2}{ak_{\rm B}T}\right)^2
              \langle \delta Z^2 \rangle \right]
              \nonumber\\
       &\ \ & \times {\rm erfc}\left[\frac{q^2}{ak_{\rm B}T}
         \sqrt{\mathstrut\frac{\langle \delta Z^2 \rangle}{2}}
               \left.  + \frac{\langle Z \rangle}
                 {\sqrt{\mathstrut 2\langle \delta Z^2 \rangle}} \right]
      \right).
\label{2511}
\end{eqnarray}


\clearpage

\end{document}